\documentstyle[12pt,aasms4]{article}

\def\arcsecpoint{$''\!.$}
\def\deg{$^{\rm o}$}

\received{}
\accepted{}

\lefthead{Crenshaw \& Kraemer}
\righthead{Kinematics in NGC 1068}

\begin{document}

\title{Resolved Spectroscopy of the Narrow-Line Region in NGC 1068:
Kinematics of the Ionized Gas\altaffilmark{1}}

\author{D. Michael Crenshaw\altaffilmark{2}
\& Steven B. Kraemer\altaffilmark{3}}
\affil{Catholic University of America and Laboratory for Astronomy and
Solar Physics, NASA's Goddard Space Flight Center, Code 681,
Greenbelt, MD  20771}

\altaffiltext{1}{Based on observations with the NASA/ESA {\it Hubble Space 
Telescope}, which is operated by the Association of Universities for Research 
in Astronomy, Inc., under NASA contract NAS5-26555.}
\altaffiltext{2}{crenshaw@buckeye.gsfc.nasa.gov}
\altaffiltext{3}{stiskraemer@yancey.gsfc.nasa.gov}

\begin{abstract}

We have determined the radial velocities of the [O~III] emitting gas in the 
inner narrow-line region (NLR) of the Seyfert 2 galaxy NGC 1068, along a slit at 
position angle 202\deg, from STIS observations at a spatial resolution of 
0\arcsecpoint1 and a spectral resolving power of $\lambda$/$\Delta\lambda$ 
$\approx$ 1000. We use these data to investigate the kinematics of the NLR 
within 6$''$ ($\sim$430 pc) of the nucleus. The emission-line knots show 
evidence for radial acceleration, to a projected angular distance of 
1\arcsecpoint7 in most cases, followed by deceleration that approaches the 
systemic velocity at a projected distance of $\sim$4$''$. We find 
that a simple kinematic model of biconical radial outflow can match the general 
trend of observed radial velocities. In this model, the emitting material is 
evacuated along the bicone axis, and the axis is inclined 5\deg\ out of the 
plane of the sky. The acceleration of the emission-line clouds provides support 
for dynamical models that invoke radiation and/or wind pressure. We suggest that 
the deceleration of the clouds is due to their collision with a patchy 
and anistropically distributed ambient medium.

\end{abstract}


\section{Introduction}

The narrow-line region (NLR) in Seyfert galaxies is characterized by emission 
lines with widths on the order of 500 km s$^{-1}$ (full-width at half-maximum, 
FWHM), which are attributed to motions of the ionized clouds of gas. Over the 
past couple of decades, ground-based studies attempted to determine the 
kinematics of the NLR, but a general consensus on the velocity flow pattern was 
not reached; cases were made for infall, rotation, parabolic orbits, outflow, 
etc. (e.g., Osterbrock and Mathews 1986, and references therein; DeRobertis \& 
Shaw 1990; Veilleux 1991; Moore \& Cohen 1994, 1996). Since the majority of the 
NLR flux comes from a region that subtends only a few arcseconds in most 
Seyferts (Schmitt \& Kinney 1996), these studies had to rely on 
spatially-integrated line 
profiles. Unfortunately, similar profile shapes and asymmetries can be 
generated from a wide variety of kinematic models (Capriotti, Foltz, \& Byard 
1980, 1981), and hence the difficulty in determining the velocity fields from 
these data. By contrast, ground-based studies of the extended narrow-line region 
(ENLR, at distances typically $\geq$ 500 pc from the central source) can take 
advantage of of spatially-resolved spectra; these studies find that the ionized 
gas in the ENLR is undergoing normal galactic rotation (Unger et al. 1987), with 
evidence for an additional component of outward radial motion in some cases 
(Whittle et al. 1988). Despite the limited spatial resolution, recent 
ground-based studies suggest that the NLR of NGC 1068, the nearest bright 
Seyfert 2 galaxy, shows evidence for radial outflow (Cecil, Bland, \& Tully 
1990; Arribas, Mediavilla, and Garc\'{i}a-Lorenzo 1996), which is a suggestion 
first offered by Walker (1968).

With the {\it Hubble Space Telescope} (HST) and the Space Telescope Imaging 
Spectrograph (STIS), we now have the ability to obtain spectra of the NLR at 
high spatial resolution. The importance of these observations is that we can 
probe the velocity field of the ionized gas close to the central continuum 
source, where the supermassive black hole presumably dominates the kinematics 
(due to its gravitational influence and/or the radiation, winds, and jets 
emanating from its vicinity). In this letter, we use STIS long-slit spectra of 
the Seyfert 2 galaxy NGC 1068 to determine the kinematics of the ionized gas in 
its NLR. In previous papers, we used these spectra to study the extended 
continuum emission (Crenshaw \& Kraemer 2000, hereafter Paper I) and the 
physical conditions in ionized gas near the continuum ``hot spot'' (Kraemer \& 
Crenshaw 2000, Paper II).
We adopt a systemic redshift of cz $=$ 1148 km~s$^{-1}$ from H~I observations 
of NGC 1068 (Brinks et al. 1997) and a distance of 14.4 Mpc (Bland-Hawthorne 
1997), so that 0\arcsecpoint1 corresponds to 7.2 pc.

\section{Observations and Results}

The observations and data reduction are described in detail in Paper I. Briefly, 
we obtained STIS low-dispersion spectra over the range 1150 -- 10,270 \AA\ at a 
spatial resolution of 0\arcsecpoint05 -- 0\arcsecpoint1 and a spectral revolving 
power of $\lambda$/$\Delta\lambda$ $\approx$ 1000 through a 0\arcsecpoint1 slit 
at a position angle of 202\deg. The slit position intercepts the brightest 
clouds in the inner NLR, as shown in Paper I. Here we concentrate on the 
brightest emission line, [O~III] $\lambda$5007, to trace the kinematics as far 
as possible away from the nucleus.

Figure 1 presents an enlarged view of the STIS spectrum, in the region around 
the H$\beta$ and [O~III] $\lambda\lambda$4959, 5007 lines, which show a 
considerable amount of spatial and velocity structure. The most striking aspect 
of these lines is that they split into two velocity components both above (SW 
of) and below (NE of) the spectrum of the continuum hot spot (the horizontal 
streak). The brightest emission-line clouds show a definite trend of increasing 
radial velocity with distance from the hot spot, out to an angular distance of 
about $\pm$1\arcsecpoint7. Most of the fainter clouds follow this trend, and 
show an overall decrease in radial velocity further out, until they approach the 
systemic velocity. There are a few exceptions to these trends, which we will 
discuss below.

We determined radial velocities from the [O~III] $\lambda$5007 emission at each 
pixel location along the slit (at a spacing of 0\arcsecpoint05). At each 
location, we fit the emission with a local continuum and a Gaussian for each 
clearly identifiable peak, resulting in 1 or 2 kinematic components. A 
number of the observed components, particularly near the hot spot, are 
asymmetric and/or very broad, and we suspect that these may split into multiple 
components at higher spectral resolution. In Figure 2, we plot the heliocentric 
radial velocities, widths (FWHM, corrected for the instrumental profile), and 
fluxes as a function of distance from the peak of the hot spot in our slit 
(which is 0\arcsecpoint14 north of the hot spot's centroid in {\it HST} optical 
continuum images, see Paper I).

Most of the radial velocities in Figure 2 follow well-defined curves; the local 
peaks in the radial velocity curves can be attributed in many cases to bulk 
motion of the emission-line knots seen in the bottom plot. The pattern of 
increasing velocity out to $\sim$1\arcsecpoint7 is evident, except for the 
blueshifted knots on the SW side, which reach a smaller maximum velocity closer 
in ($\sim$1$''$). The decrease in radial velocity at greater distances is 
also clear, except 
that the redshifted knots in the SW abruptly terminate at about 2\arcsecpoint0. 
There are two knots of emission NE of the hot spot that do not 
conform to this pattern at all; one is highly blueshifted ($-$1400 km s$^{-1}$) 
and the other is highly redshifted ($+$1000 km s$^{-1}$) with respect to the 
systemic velocity. The middle plot in Figure 2 shows the large widths of the 
lines, which tend to decrease with distance, particularly in the SW.

\section{A Kinematic Model: Biconical Outflow}

HST images indicate a biconical geometry for the NLR in most Seyfert 2 galaxies 
(Schmitt \& Kinney 1996), and for NGC 1068 in particular (Evans et al. 1991). 
The radial velocity curves in Figure 2 suggest a velocity field in which the 
emission-line knots accelerate out from the inner nucleus, reach a terminal 
velocity, and then decelerate. Thus, we favor a kinematic model of biconical 
outflow away from the nucleus. Similar amplitudes of the blueshifted and 
redshifted curves on the NE side indicate that the axis of the bicone is close 
to the plane of the sky, and the lack of low radial velocities where the curves 
peak (around $\pm$1\arcsecpoint7) suggests that the bicone is evacuated along 
its axis.

We have generated a kinematic model of radial outflow in a bicone that is hollow 
along its axis. We constrain our model to be consistent with the observed 
morphology: Evans et al. (1991) find that the NLR in NGC 1068 can be described 
by a bicone with a projected half-opening angle of $\sim$35\deg\ and a position 
angle of the bicone axis on the sky of 15\deg. For simplicity, we assume that 
the two cones have identical properties (geometry, velocity law, etc.), a 
filling factor of one within the 
defined geometry, and no absorption of 
[O~III] photons (e.g., by dust). The parameters that are allowed to vary in our 
code are the extent of each cone along its axis (z$_{max}$), its minimum and 
maximum half-opening angles ($\theta$$_{inner}$ and $\theta$$_{outer}$), the 
inclination 
of its axis out of the plane of the sky (i$_{axis}$), and the velocity law as a 
function of distance from the nucleus. For the latter, we will show that 
constant acceleration to a maximum velocity (v$_{max}$) at a turnover radius 
(r$_t$), followed by a constant deceleration to zero velocity at the greatest 
extent of the cone ($=$ z$_{max}$/cos $\theta$$_{outer}$), provide a reasonable 
match to the observations.  

Our code generates a two-dimensional velocity map and samples this map with a 
slit that matches the position, orientation, and width of our observational slit 
(Paper I); in this case, our slit is placed 0\arcsecpoint14 north of the hot 
spot centroid and rotated 7\deg\ with respect to the 
projected bicone axis. We then compare the simulated and observed radial 
velocity curves, and adjust 
the model parameters until a good match is obtained. Since the observed radial 
velocity curves have significant intrinsic scatter, we do not fine-tune the 
models (e.g., by choosing different turnover locations and maximum velocities 
for each side), but settle for an illustrative model that matches the overall 
trend. The parameters for our final model are given in Table 1.

Figure 3 shows the envelope of radial velocities from the model, compared to the 
observed radial velocities; the width of the envelope is determined by the range 
in half-opening angle ($\theta$$_{inner}$ to $\theta$$_{outer}$), and the 
relative 
amplitudes of blue and redshifted curves are determined by the inclination angle 
(i$_{axis}$). In three quadrants of the plot in Figure 3, the overall trend of 
observed radial velocities is well matched by the model.
\footnote{Note that we have assumed that the origin of the outflow is the 
continuum hot spot; if the origin is at the S1 radio component (Gallimore, Baum, 
\& O'Dea 1996), which is 0\arcsecpoint17 S of the hot spot (Capetti, Macchetto, 
\& Lattanzi 1997), we find that only minor adjustments are needed in the 
half-opening angles and turnover radius to match the oberved trend.}
For the blueshifted emission on the SW side, the radial velocities reach a 
smaller maximum closer in (at $\sim$1$''$), and show a slight trend towards 
deceleration further out. As mentioned previously, there are two emission-line 
knots at very high velocities that do not fit this model at all. The presence of 
additional points outside of the envelope indicates that there may be a few 
emission-line knots close to the axis or knots that undergo slightly different 
accelerations or decelerations. Thus, although this model is simplistic, we 
adopt it as a tool for interpreting the velocity field, and discuss ways in 
which the descrepancies can be accommodated.

\section{Discussion}

\subsection{Comparison with Ground-based Observations}

Cecil et al. (1990) provide the most comprehensive set of velocity maps for the 
NLR of NGC 1068, based on Fabry-Perot observations of the [N II] lines at 
$\sim$1$''$ spatial and 140 km s$^{-1}$ spectral resolutions; these authors 
conclude that their data can be explained by cylindrical or biconical outflow. 
Given the limited spatial resolution, their observations are in agreement with 
ours and are consistent with our model. Their line profiles along the STIS slit 
position show a single broad component at the location of the hot spot and 
double-peaked profiles, separated by $\sim$1000 km s$^{-1}$, at a distance 
$\sim$2$''$ from the hot spot, in agreement with our observations (Figure 3). 
Outside of the STIS slit position, their profiles show the same double-peaked 
structure, with the largest velocity separation at $\sim$2$''$ on either side of 
the hot spot, which is consistent with our model prediction that the highest 
blueshifts and redshifts should occur at this distance. The higher spatial 
resolution of the STIS observations allows us to see the acceleration of clouds 
outward from the nucleus followed by a clear deceleration of the clouds. We have 
not applied our model to ground-based observations at distances $>$ 6$''$, where 
galactic rotation begins to play a much larger role in the kinematics (Baldwin, 
Whittle, \& Wilson 1987).

\subsection{Comparison with Other Models}

Two other kinematic models of the NLR have been proposed on the basis of 
spectra at high spatial resolution, which were obtained with {\it HST}'s Faint 
Object Camera (FOC). Axon et al. (1998) suggest a model for NGC 1068 in which 
the gas expands outward from the radio jet (which is nearly coincident with the 
axis of the ionization cone). Winge et al. (1999) propose a rotating disk model 
for the NLR in the Seyfert 1 galaxy NGC~4151.
To test the Axon et al. (1998) suggestion for NGC 1068, we generated a model 
with the same parameters as in Table 1, except that the velocity vectors are 
directed perpendicular to the radio axis. In this case, we find that the 
envelope of radial velocities is nearly the same as in Figure 3, due to the 
small inclination angle. However, we have two concerns about this model. First, 
the observed radial velocities follow a well-organized flow pattern, and we can 
discern no correlation with the clumpy radio structure in the NLR (Gallimore et 
al. 1996). 
Second, this model cannot explain the kinematics of the NLR in the Seyfert 1 
galaxy NGC 4151, where the gas is blueshifted in one cone and redshifted in the 
other (Winge et al. 1999; Nelson et. al. 2000, and references therein), whereas 
motions perpendicular to the axis produce equal blueshifts and redshifts in each 
cone, regardless of the inclination of the axis. We note that our radial outflow 
model for NGC 1068 provides a slightly better fit than the Axon et al. model, 
because a small tilt of the axis (5\deg) can match the slightly different 
amplitudes of the observed blueshifted and redshifted curves. Furthermore, by 
tilting the cone axis towards the line of sight, this model can explain the 
observed radial velocities in NGC 4151 (see Crenshaw et al. 2000).

Winge et al.'s (1999) rotating disk model for NGC 4151 is only used to match 
their low velocity component (within 300 km s$^{-1}$ of systemic). Even so, they
require an extended (and otherwise undetected) distribution of matter within 
0\arcsecpoint1 (64 pc) of the nucleus with a mass on the order of 10$^{9}$ 
M$\odot$, in addition to a central point-source mass of 10$^{7}$M$\odot$. A 
rotation model can be ruled out for NGC 1068, because high redshifts and 
blueshifts are seen on either side of the nucleus. Other gravitational models 
can also be ruled out as the principal source of the velocities in NGC 1068, 
because the required mass is prohibitive. At the peaks of the velocity curves 
(ignoring the two knots with very high velocities), the projected distance from 
the nucleus is $\sim$120 pc, the projected velocity is $\sim$850 km s$^{-1}$, 
and the required mass is $\geq$ 10$^{10}$ M$\odot$ (a lower limit because of 
projection effects). By comparison, the dynamical mass from stars within a 
radius of $\sim$1$''$ from the nucleus of NGC 1068 is only 6 x 10$^{8}$ M$\odot$ 
(Thatte et al. 1997). Thus, radial outflow provides the simplest and best 
explanation of the observed velocities in these two Seyfert galaxies.

\subsection{Implications of our Model}

Our kinematic model assumes constant acceleration of clouds in the 
inner NLR ($<$ 140 pc), and constant deceleration further out; this assumption 
provides a reasonable match to the observations, although more complicated 
velocity laws as a function of distance are possible, given the intrinsic 
scatter in the observed points. Nevertheless, these results favor dynamical 
models that invoke radiation and/or wind pressure to drive clouds out from the 
nucleus. The deceleration of clouds further is not primarily due to gravity, for 
the reasons given above: an unreasonably high mass ($\sim$10$^{10}$ M$\odot$) is 
required to slow the clouds down. The simplest explanation for the deceleration 
is that the clouds experience a drag force, presumably due to interaction with 
ambient material at $\sim$140 pc from the nucleus. A possible explanation for 
the blueshifted clouds on the SW side is that they run into ambient material 
that is closer to the nucleus ($\sim$80 pc). In this picture,
the two high velocity clumps in Figure 3 represent clouds that have not 
experienced a drag force in the direction they are traveling, which suggests 
that there are holes in the surrounding medium. 
We note that gravity may eventually play a role in the kinematics. In our model, 
the axis of the outflow is inclined by $\sim$45\deg\ with respect to the 
galactic disk (inclination $=$ 40\deg, major axis at position angle 106 \deg, 
see Bland-Hawthorne et al. 1997), and as the NLR clouds slow down, they may be 
pulled back to the disk, possibly joining the existing ENLR gas.

In conclusion, we find that a biconical outflow model, with evidence for 
acceleration close to the nucleus and deceleration further out, provides a 
reasonable explanation for the radial velocities in our long-slit spectrum of 
NGC 1068. STIS long-slit observations of NGC 1068 at higher spectral resolution 
and at different slit positions will be helpful in resolving velocity 
components and mapping out the two-dimensional velocity field. STIS observations 
of other Seyferts will help test the utility of this model.

\acknowledgments
This work was supported by NASA Guaranteed Time Observer funding to the STIS 
Science Team under NASA grant NAG 5-4103.

\clearpage

\figcaption[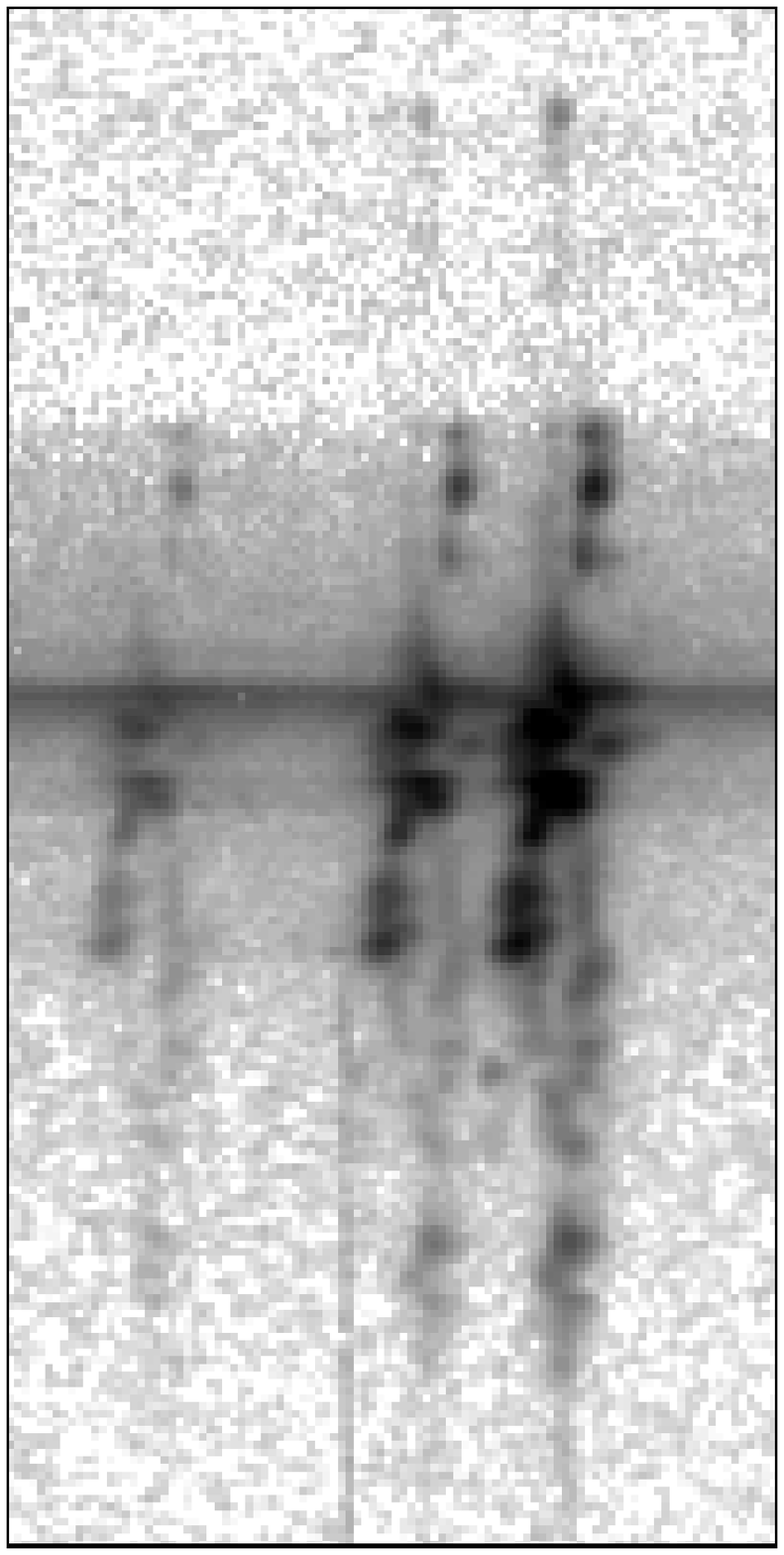]{Expanded view of the STIS G430L spectral image of NGC 1068, 
showing the H$\beta$ and [O~III] $\lambda\lambda$4959, 5007 lines (wavelength 
increases to the right). The slit position angle is 202\deg; NE is below and SW 
is above the spectrum of the hot spot (the bright continuum spectrum). (The 
bright column to the left of the [O~III] $\lambda$4959 line is a bad CCD 
column.) The spatial extent of this figure is 10$''$.}

\figcaption[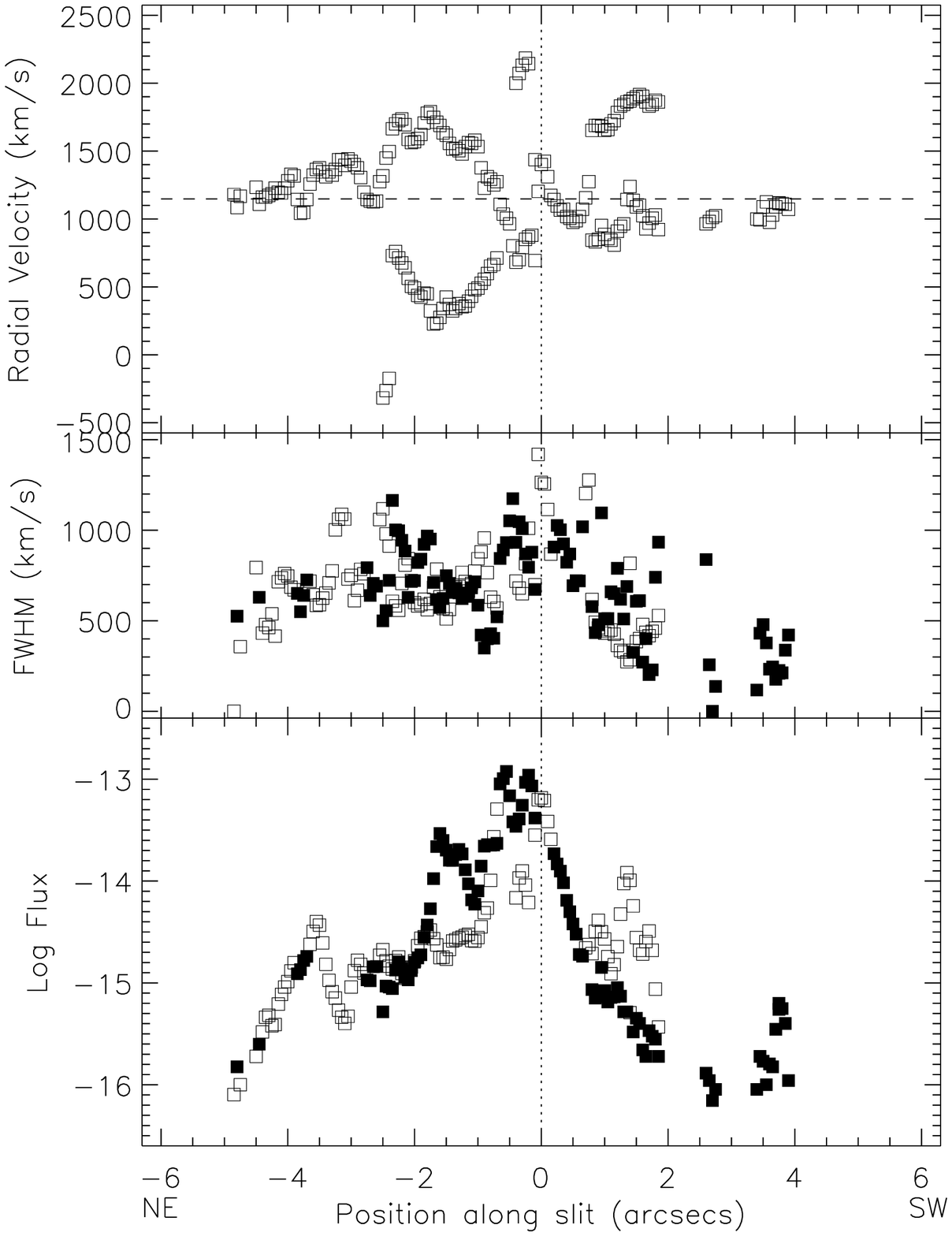]{Radial velocity, width (FWHM, corrected for instrumental 
width), and log flux (ergs s$^{-1}$ cm$^{-2}$ per cross-dispersion pixel) of the 
[O~III] $\lambda$5007 emission along the slit, as a function of angular distance 
from the peak of the hot spot spectrum. Negative positions are towards the NE. 
In the bottom two plots, filled and open squares represent blueshifted and 
redshifted emission, respectively (relative to systemic).} 

\figcaption[fig23ps]{Comparison of observed and model radial velocities. Errors 
in the observed radial velocities are smaller than the symbols, except as noted. 
The shaded region is the envelope of predicted radial velocities from the 
kinematic model described in the text.}

\clearpage
\begin{deluxetable}{ll}
\tablecolumns{2}
\footnotesize
\tablecaption{Kinematic Model of NGC 1068\label{tbl-1}}
\tablewidth{0pt}
\tablehead{
\colhead{Parameter} & \colhead{Value}
}
\startdata
z$_{max}$         &306 pc \\
$\theta$$_{inner}$  &26\deg\                  \\
$\theta$$_{outer}$  &40\deg\                  \\
i$_{axis}$        &5\deg\ (NE is closer)    \\
v$_{max}$         &1300 km s$^{-1}$         \\
r$_t$             &137 pc \\
\enddata
\end{deluxetable}



\clearpage
\vskip3.0in
\begin{figure}
\plotone{fig1.ps}
\\Fig.~1.
\end{figure}

\clearpage
\begin{figure}
\plotone{fig2.ps}
\\Fig.~2.
\end{figure}

\clearpage
\begin{figure}
\plotone{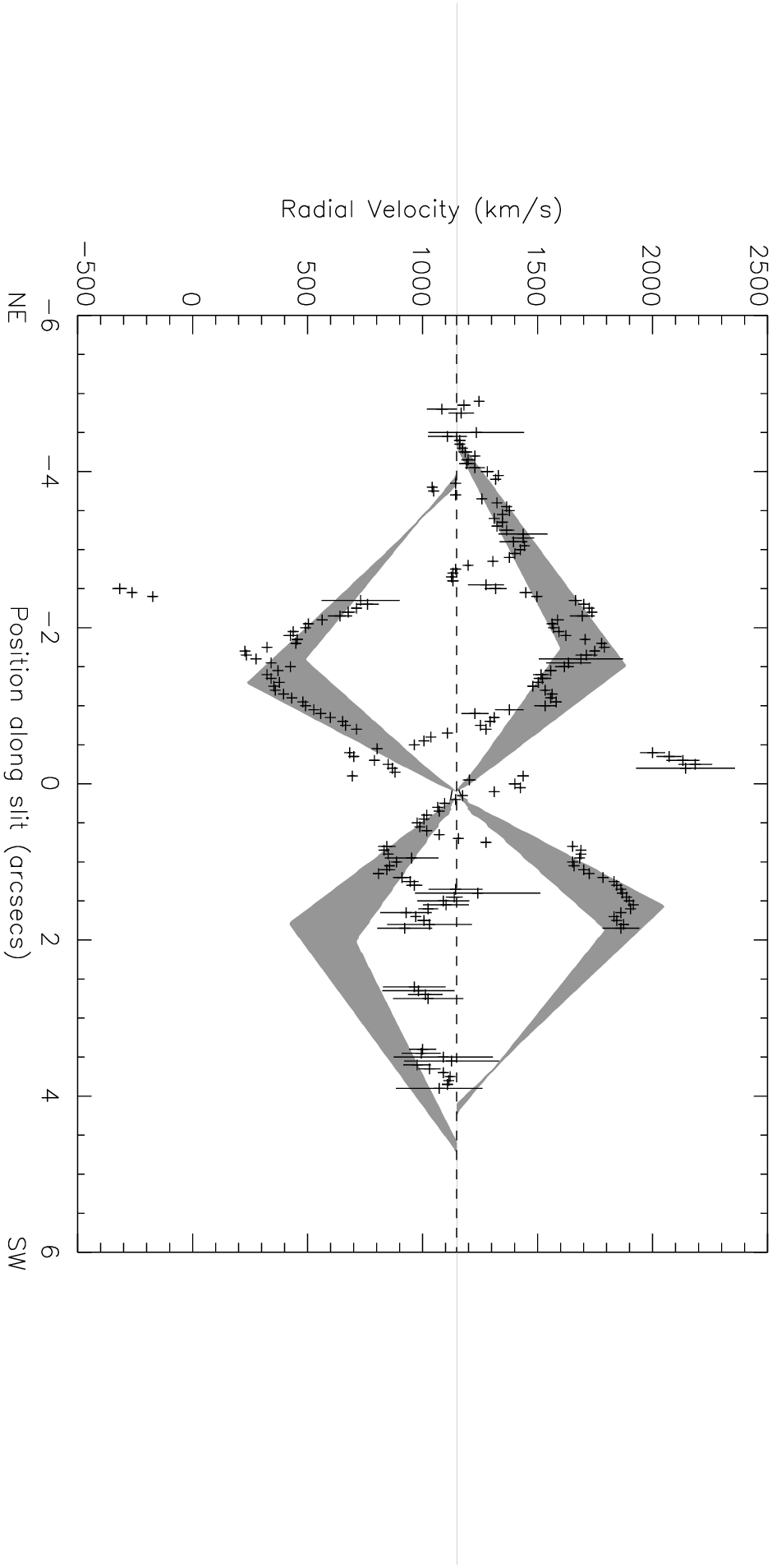}
\\Fig.~3.
\end{figure}

\end{document}